\documentclass[10pt,journal,compsoc]{IEEEtran}
\usepackage{ragged2e}
\newif\ifpeerreview

\peerreviewfalse

\usepackage[nocompress]{cite}
\usepackage{url}
\usepackage{amsmath,amssymb,graphicx}

\usepackage{lipsum} 
\usepackage{multirow}
\usepackage{booktabs}
\usepackage{xcolor}
\usepackage[switch]{lineno}
\usepackage{hyperref}
\usepackage{arydshln}
\newcommand{\paperID}{28}

\title{Q-Agent: Quality-Driven Chain-of-Thought 
Image Restoration Agent through Robust Multimodal Large Language Model}

\author{Yingjie~Zhou \textsuperscript{1},
        Jiezhang~Cao \textsuperscript{1},
        Farong Wen \textsuperscript{1},
        Zicheng Zhang \textsuperscript{3},
        Yu Zhou \textsuperscript{1},
        Yue Shi \textsuperscript{3},
        Xiaohong Liu \textsuperscript{1,2},
        Radu Timofte \textsuperscript{4},
        Luc Van Gool \textsuperscript{5},
        and Guangtao Zhai \textsuperscript{1},~\IEEEmembership{Fellow,~IEEE}
\IEEEcompsocitemizethanks{\IEEEcompsocthanksitem \textsuperscript{1} Shanghai Jiao Tong University
\IEEEcompsocthanksitem \textsuperscript{2} Suzhou Key Laboratory of Artificial Intelligence
\IEEEcompsocthanksitem \textsuperscript{3} Shanghai Artificial Intelligence Laboratory
\IEEEcompsocthanksitem \textsuperscript{4} University of Würzburg
\IEEEcompsocthanksitem \textsuperscript{5} Institute for Computer Science, Artificial Intelligence and Technology}
}

\begin{document}

\IEEEtitleabstractindextext{%
\begin{abstract} Image restoration (IR) often faces various complex and unknown degradations in real-world scenarios, such as noise, blurring, compression artifacts, and low resolution, etc. Training specific models for specific degradation may lead to poor generalization. 
To handle multiple degradations simultaneously, All-in-One models might sacrifice performance on certain types of degradation and still struggle with unseen degradations during training. Existing IR agents rely on multimodal large language models (MLLM) and a time-consuming rolling-back selection strategy neglecting image quality.
As a result, they may misinterpret degradations and have high time and computational costs to conduct unnecessary IR tasks with redundant order. 
To address these, we propose a Quality-Driven agent (Q-Agent) via Chain-of-Thought (CoT) restoration.
Specifically, our Q-Agent consists of robust degradation perception and quality-driven greedy restoration.
The former module first fine-tunes MLLM, and uses CoT to decompose multi-degradation perception into single-degradation perception tasks to enhance the perception of MLLMs. The latter employs objective image quality assessment (IQA) metrics to determine the optimal restoration sequence and execute the corresponding restoration algorithms. Experimental results demonstrate that our Q-Agent \textbf{achieves superior IR performance} compared to existing All-in-One models. 
\end{abstract}

}

\ifpeerreview
\linenumbers \linenumbersep 15pt\relax 
\author{Paper ID \paperID\IEEEcompsocitemizethanks{\IEEEcompsocthanksitem This paper is under review for ICCP 2026 and the PAMI special issue on computational photography. Do not distribute.}}
\markboth{Anonymous ICCP 2026 submission ID \paperID}%
{}
\fi
\maketitle

\IEEEraisesectionheading{
  \section{Introduction}\label{sec:introduction}
}
%
%
%
%
\IEEEPARstart{I}mage restoration (IR) aims to recover distorted images and enhance visual quality. Over the years, Task-Specific IR methods \cite{scunet,wu2023ridcp,Cai_2023_ICCV,zhou2024reli} have addressed specific types of IR. However, these methods often have poor generalization on 
real-world distortions. To address this issue, All-in-One IR frameworks restore multiple types of degradation within a unified approach. While these methods improve generality, studies \cite{xiang2023deep,chen2025restoreagent} have shown that they often fall short in effectiveness and flexibility compared to Task-Specific models. Consequently, balancing effectiveness and generality in IR remains a critical and ongoing research challenge. 

\begin{figure}[!t]
    \vspace{-0cm}
    \centering
    \includegraphics[width =1\linewidth]{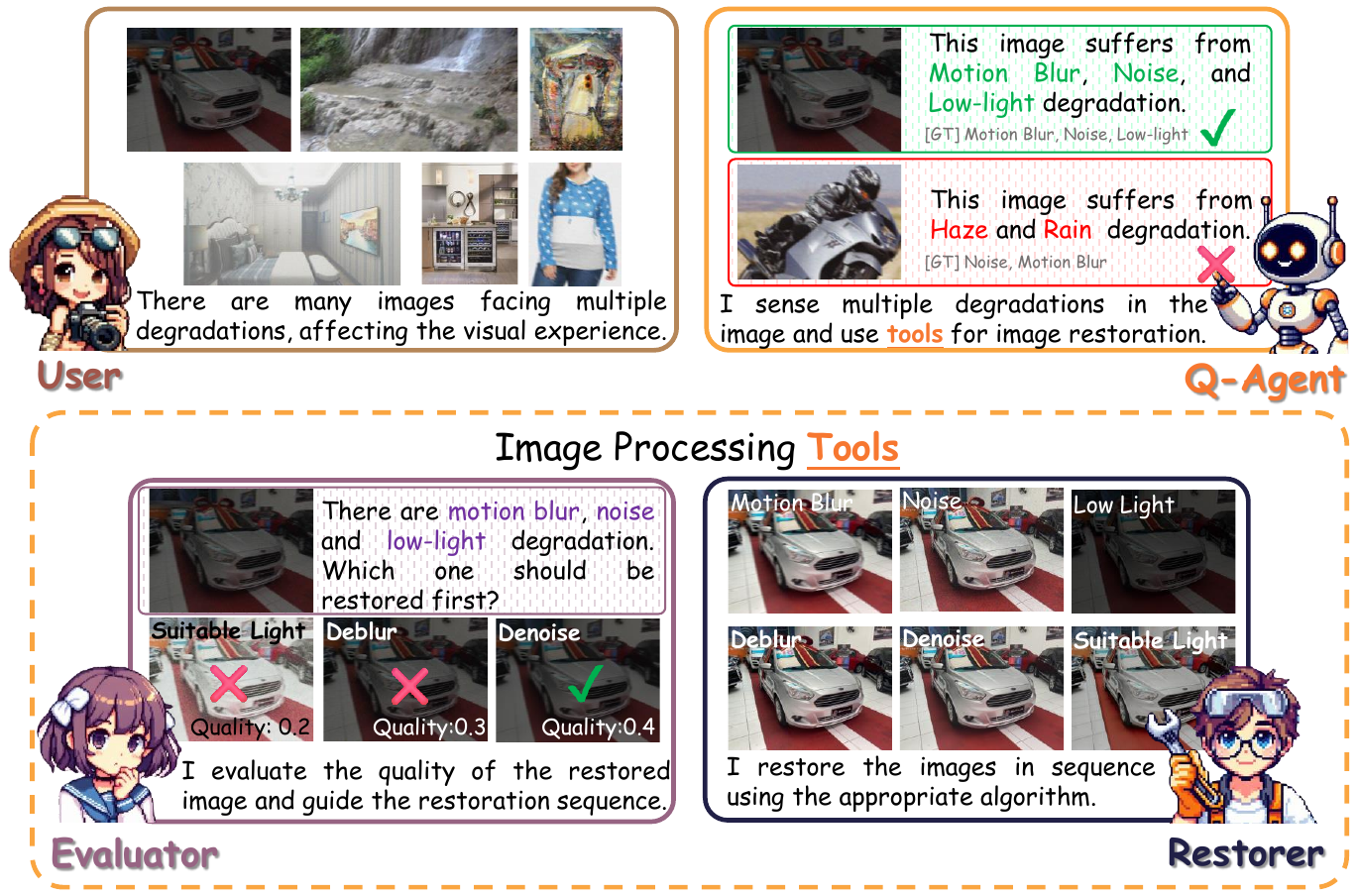}
    \vspace{-0.7cm}
    \caption{Q-Agent design motivation. Existing MLLMs and image processing tools provide new solutions for realizing multiple degradation IR.}
    \label{fig:teaser}
    \vspace{-0.6cm}
\end{figure}

The rise of Multimodal Large Language Models (MLLMs) presents new opportunities for IR, as these models demonstrate strong capabilities in multitask scheduling and processing, as shown in Fig.~\ref{fig:teaser}. 
However, MLLMs still suffer from hallucinations \cite{huang2024opera}, which can reduce the reliability of their outputs. This issue becomes particularly critical in low-level visual perception tasks, where incorrect degradation perception can directly mislead subsequent image enhancement and processing. As highlighted by Zhang $et$ $al.$ \cite{zhang2024q}, MLLMs exhibit limitations in accurately perceiving degraded images, especially when multiple types of degradation are present. This poses a significant challenge in the design of IR agents based on MLLMs. Furthermore, determining the optimal IR sequence remains a crucial problem. While Chen $et$ $al.$ \cite{chen2025restoreagent} proposed a rollback-based approach to establish the restoration order, its computational complexity increases significantly with the number of degradation types, leading to inefficiencies and scalability constraints. Therefore, developing a more efficient restoration sequence selection method remains challenging.

To address the aforementioned challenges, we propose Q-Agent, a quality-driven IR agent that integrates robust degradation perception with quality-driven greedy restoration, enabling a more intelligent and interactive end-to-end restoration process. To enhance the degradation perception capability of MLLMs, we construct Q-Degrade, a large-scale dataset containing images with eight distinct levels of degradation and the sequential order in which these degradations are introduced. By combining fine-tuning techniques with chain-of-thought (CoT) \cite{wei2022chain} reasoning, we enable MLLMs to achieve the more robust and accurate perception of multiple degradation types. Additionally, we introduce five no-reference (NR) image quality assessment metrics and different IR algorithms, which serve as the core tools for Q-Agent. To optimize the restoration sequence, we propose a quality-driven greedy algorithm, which leverages objective NR quality assessment metrics to iteratively select the restoration step that maximizes quality improvement. This approach ensures an optimal restoration sequence, significantly reducing computational overhead compared to rollback-based methods. Experimental results demonstrate that Q-Agent achieves restoration performance exceeding existing All-in-One methods, while maintaining high flexibility and scalability. In conclusion, the main contributions of this paper are as follows:

\begin{itemize} 
\item We propose a robust degradation perception method based on CoT. It decomposes the multiple-degradation perception problem into multiple subproblems with separate degradation perceptions, thus allowing our MLLM to obtain better perception than existing IR agents.
\vspace{1mm}
\item We propose a quality-driven greedy strategy, which measures the image quality to guide the determination of greedy restoration order, effectively avoiding the number of rollbacks and improving the computational efficiency and scalability of the framework. Importantly, our algorithm has a linear time complexity, which is significantly lower than the complexity of existing IR agents. 
\vspace{1mm}
\item Our Q-Agent achieves optimal performance and is highly flexible and extensible compared to existing All-in-One models and is expected to be expanded into a more powerful and comprehensive IR agent.

\end{itemize}

\section{Related Works}
\label{sec:formatting}
\vspace{-5pt}
\subsection{Image Restoration}
Images often suffer from degradations, such as noise, blur, JPEG compression artifacts, low illumination and resolution.
To preserve the image quality, targeted restoration of degraded images on commonly used datasets \cite{kodak,yang2017deep,nah2017deep,abuolaim2020defocus,wang2021seeing,rim2020real,mei2024mssidd,shen2019human} is essential. 
However, these datasets remain limited in scale and diversity of distortions, thereby restricting their ability to comprehensively model the degradations encountered in real-world scenarios. To mitigate these limitations, we introduce Q-Degrade, a large-scale image degradation dataset comprising 100,000 degraded images, aimed at advancing research in this field. 

The evolution of IR methods can be categorized into two paradigms: Task-Specific \cite{scunet,wu2023ridcp,Cai_2023_ICCV,zhou2024reli} and All-in-One models \cite{realesrgan,stablesr,promptir,daclip,airnet,mioir,autodir,insturctir}. Task-Specific models are designed to effectively restore images affected by a particular type of degradation; however, they exhibit notable limitations when addressing multiple degradation types simultaneously. In contrast, All-in-One models adopt a unified architecture capable of handling multiple degradations within a single framework. While this approach enhances generalization, it often comes at the cost of restoration performance. To balance restoration effectiveness and generalizability, this paper proposes Q-Agent, a novel IR framework based on MLLMs. Q-Agent leverages the perceptual capabilities of MLLMs to comprehensively assess image degradation in an All-in-One manner. By intelligently selecting and sequentially applying appropriate IR tools, Q-Agent enhances the visual quality of degraded images, offering a more adaptive and effective solution to IR.

\subsection{Image Quality Assessment (IQA)}
IQA aims to evaluate both the visual quality of images and the effectiveness of IR methods. In general, IQA can be categorized into subjective and objective quality assessments. Subjective quality assessment involves human observers who evaluate image quality based on visual perception, providing the most accurate reflection of human vision. However, this approach is limited in practical applications due to its high cost and time-consuming nature, requiring the recruitment of participants and controlled experimental conditions. Conversely, objective quality assessment employs algorithms to estimate image quality, making it more practical and widely adopted. Objective IQA methods are further classified into Full-Reference (FR) \cite{dhhqa}, Reduced-Reference (RR) \cite{zhang2024reduced}, and No-Reference (NR) \cite{thqa3d,zhou2023no} approaches. FR IQA metrics, such as PSNR, SSIM \cite{ssim} and LPIPS \cite{lpips}, assess image quality by comparing a degraded or restored image with its pristine counterpart. In contrast, NR IQA metrics (e.g., BRISQUE \cite{brisque}, NIQE \cite{niqe}, and CPBD \cite{cpbd}) estimate image quality solely from the degraded image, without requiring a reference. The RR IQA method serves as an intermediate approach, utilizing partial information from the reference image to facilitate quality assessment.

In the domain of IR, FR IQA metrics are the most widely used, as they provide a direct and quantifiable measurement of restoration effectiveness by computing differences between the original and restored images. However, in many real-world applications, high-quality reference images are often unavailable, rendering FR IQA methods impractical. In such cases, NR IQA methods offer a more flexible and scalable solution, enabling efficient and autonomous IQA without reliance on a reference image.

\section{Dataset Construction}

\subsection{Source Image Collection}
To ensure the highest quality and most diverse image dataset, we select 12,225 images from the Laion-High-Resolution dataset \cite{schuhmann2022laion} as source images. The selection encompasses a broad range of categories, including both real-world photographs and computer-generated images. In terms of content, the dataset includes a diverse array of subjects such as characters, scenes, objects, artistic paintings and advertisements. To provide a more intuitive representation of this diversity, we have included visualizations of a subset of the selected images in Supplementary Material.

\subsection{Multiple Degradation Simulation}
\begin{figure}[!t]
    \vspace{-0cm}
    \centering
    \includegraphics[width =1\linewidth]{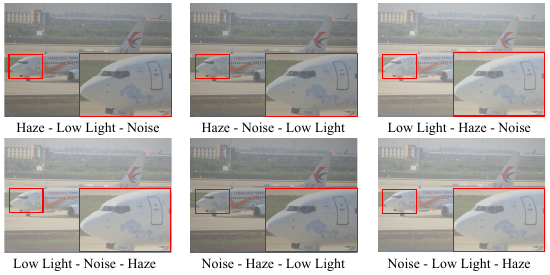}
    \vspace{-0.7cm}
    \caption{Different effects of different degradation addition orders on the synthesized degradation images.}
    \label{fig:addseq}
    \vspace{-0.5cm}
\end{figure}

Eight common types of degradation are selected for this study: noise, motion blur, out-of-focus blur, JPEG compression artifacts, low light, low resolution, haze, and rain. These degradations are simulated using computational methods, with their respective parameters carefully controlled. A more detailed description of the degradation simulation is provided in the Supplementary Material. While individual degradation types can guide the development of IR algorithms, such models tend to be overly simplistic. In real-world scenarios, images are often affected by a combination of degradation types, leading to more complex and integrated forms of image distortion. To better simulate these real-world conditions, the Q-Degrade dataset is constructed by randomly applying up to four degradation types, selected from the eight listed above, to the source images. This approach aims to mimic the multiple, concurrent degradations that are typically encountered in practical image acquisition. Notably, unlike prior works \cite{chen2025restoreagent,ddhqa,sjtuh3d}, the selected degradation types are applied in a specific order, a decision motivated by two factors: 1) Although the degradation sequence cannot be directly observed in the image, it exists objectively throughout the image degradation process; 2) As illustrated in Fig.~\ref{fig:addseq}, the order of applied degradations can significantly influence the visual effects of the final image. Thus, the degradation sequence is an important consideration in the dataset construction process. Taking into account various factors, each source image undergoes ten distinct instances of degradation, with variations in the type, severity, and application order of the distortions. This procedure results in a total of 125,550 degraded images. After manual inspection, a number of extreme cases are removed, including excessively degraded images and those containing sensitive content. The final version of the Q-Degrade dataset consists of 100,000 degraded images.

\begin{figure*}[!t]
    \vspace{-0cm}
    \centering
    \includegraphics[width =1\linewidth]{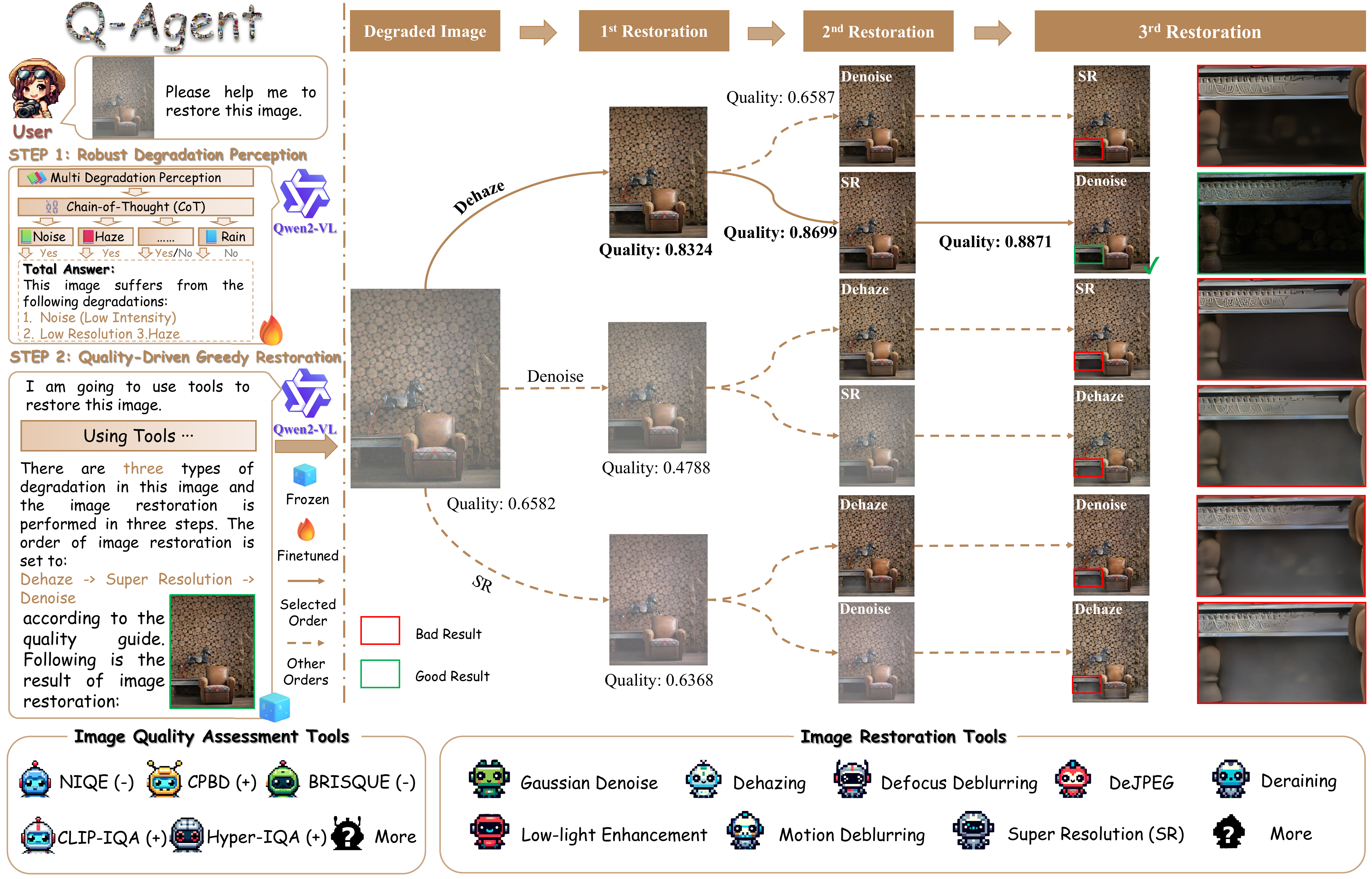}
    \vspace{-0.6cm}
    \caption{The framework of Q-Agent, which is based on Qwen2-VL \cite{qwen2vl}, enables end-to-end image restoration by using various image quality assessment tools and image restoration tools. The plus and minus signs of the IQA tools indicate the correlation between the predicted values and the image quality. Additional tools can be integrated to further enhance Q-Agent's image restoration capabilities. }
    \label{fig:framework}
    \vspace{-0.4cm}
\end{figure*}

         

\begin{figure}[!t]
    \vspace{-0cm}
    \centering
    \includegraphics[width =1\linewidth]{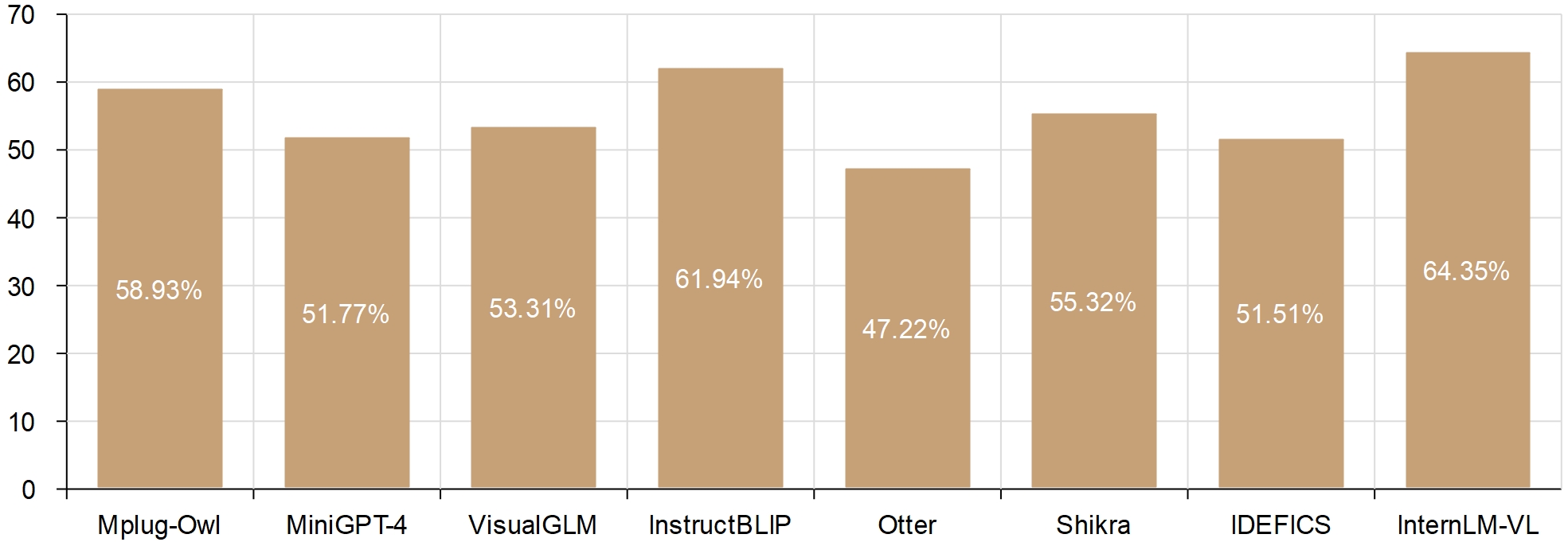}
    \caption{Degradation-aware accuracy of MLLMs on Q-Bench.}
    \label{fig:bench}
    \vspace{-0.6cm}
\end{figure}

\section{Q-Agent for CoT Restoration}
The high computational complexity is a serious challenge for existing IR agent \cite{chen2025restoreagent}, especially in terms of sequence determination for multi-degraded IR, which limits further expansion and optimization.
In this paper, we propose a new restoration agent (called of Q-Agent) via CoT \cite{wei2022chain}, and its framework is shown in Fig.~\ref{fig:framework}.
Our Q-Agent contains two modules: a robust degradation perception module and a quality-driven greedy restoration module.
The former combines fine-tuning and CoT \cite{wei2022chain} techniques to enhance the multiple degradation perception of MLLM to formulate accurate restoration tasks. The latter, on the other hand, realizes the determination of optimal restoration order and IR guided by image quality.

\subsection{Robust Degradation Perception}
One can directly apply MLLM in an agent for the IR task with multiple degradations. 
However, it may lead to inevitable issues because MLLM has inherent limitations on degradation prediction.
MLLM might result in incorrect degradation perception and further obtain misleading sequence of multiple single IR tasks.
As shown in Fig. \ref{fig:bench}, the Q-Bench proposed by Zhang $et$ $al.$ \cite{wu2023q} provides a benchmark for comparing the degradation-awareness capabilities of various MLLMs \cite{li2023otter,ye2023mplug,dai2023instructblipgeneralpurposevisionlanguagemodels,chen2023shikra,glm2024chatglm,zhu2023minigpt,laurencon2023obelics,chen2024internvl}, from which it can be seen that the performance of most MLLMs in dealing with image degradation related problems is far from satisfying the requirement of accurately performing multiple degradation perception.

Image degradation perception involves recognizing the various types of degradation and their respective degrees in degraded images, which is crucial for accurate and efficient IR. In order to improve the perception of MLLM on multiple degraded images, we first fine-tune MLLM using Low-Rank Adaptation (LoRA) \cite{hu2022lora}. 

In addition, we use CoT \cite{wei2022chain} to decompose the complex multiple degradation perception task into simple \textit{Yes-or-No} questions for individual degradation types. Specifically, MLLM is asked one by one \textit{``Is there \textless dis\textgreater\ in this image?"}, where \textit{\textless dis\textgreater}\ denotes various degradation types. In particular, when \textit{\textless dis\textgreater}\ is \textit{``noise”} and the response is \textit{``Yes,"} MLLM is further questioned to determine the degree of degradation with the request \textit{``What is the intensity of the noise present in this image: A. low B. medium C. high."} When all degradation types have been questioned, the degradation types that responded \textit{``Yes"} are used as the result to develop subsequent restoration tasks.

\subsection{Quality-Driven Greedy Restoration}
\begin{figure}[!t]
    \vspace{-0cm}
    \centering
    \includegraphics[width =1\linewidth]{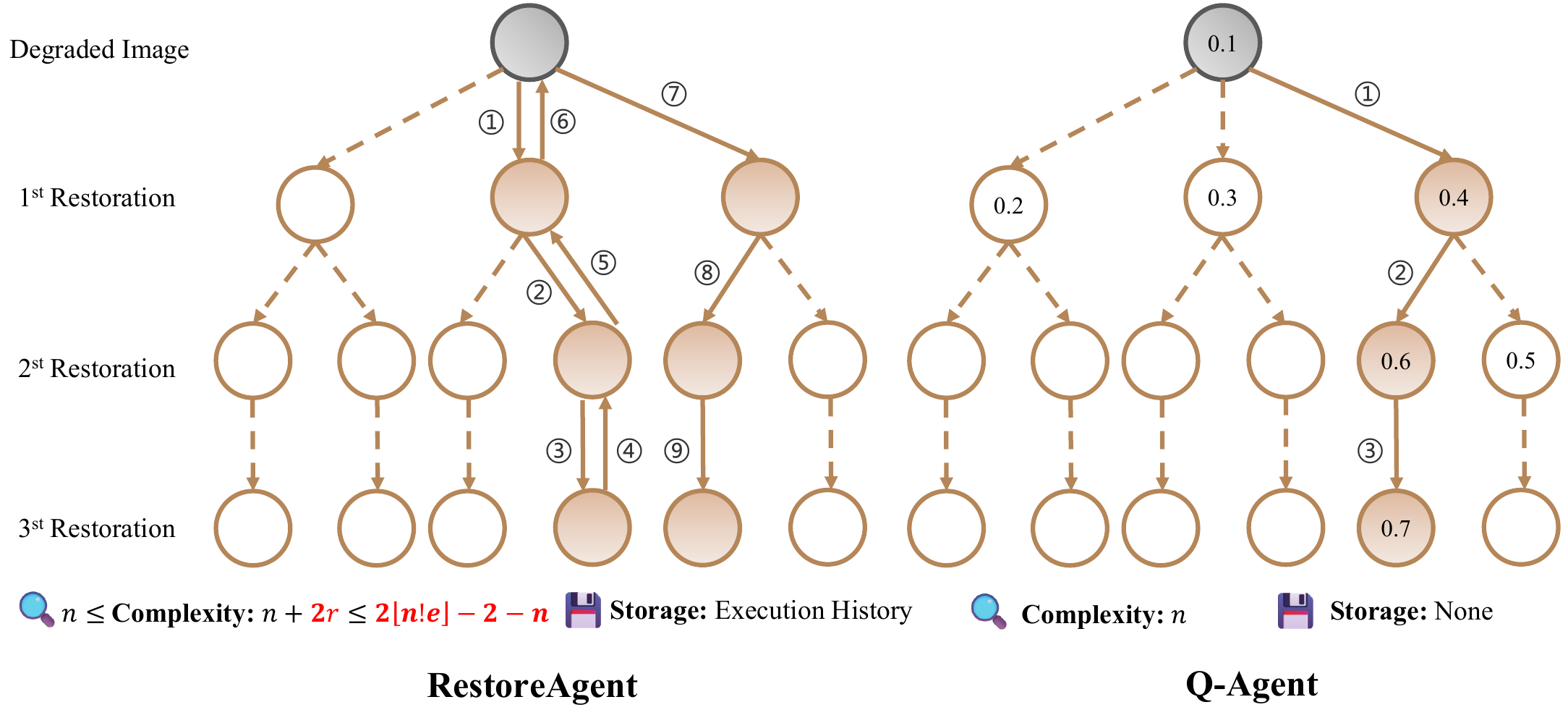}
    \caption{The superiority of Q-Agent. Assume that the number of restoration stages and rollbacks is $n$ and $r$. RestoreAgent's worst-case complexity, derived in the Supplementary Material, highlights the limitations in dealing with more degradations.}
    \label{fig:search}
    \vspace{-0.6cm}
\end{figure}

Images often suffer multiple degradations in real-world scenarios.
Removing these degradations in different orders significantly impacts the quality of the restored images.
Thus, designing an optimal restoration order represents a critical challenge in developing an IR agent. 
Existing agent methods based on rolling back \cite{chen2025restoreagent} and reinforcement learning (RL) \cite{chen2025restoreagent,ning2024survey,rathbun2025sleepernets} may lead to high computational costs.
RestoreAgent \cite{chen2025restoreagent} uses rolling-back operations to identify the optimal restoration order. However, Fig.~\ref{fig:search} reveals that this approach incurs significant computational and storage overhead. 
Unlike the RL-based agents, we aim to design an agent that does not require additional training. 
In this paper, we propose Q-Agent by employing a quality-driven greedy strategy to select the restoration order. 

\subsubsection{Quality-Driven Greedy Strategy}
\begin{figure}[!t]
    \vspace{-0cm}
    \centering
    \includegraphics[width =1\linewidth]{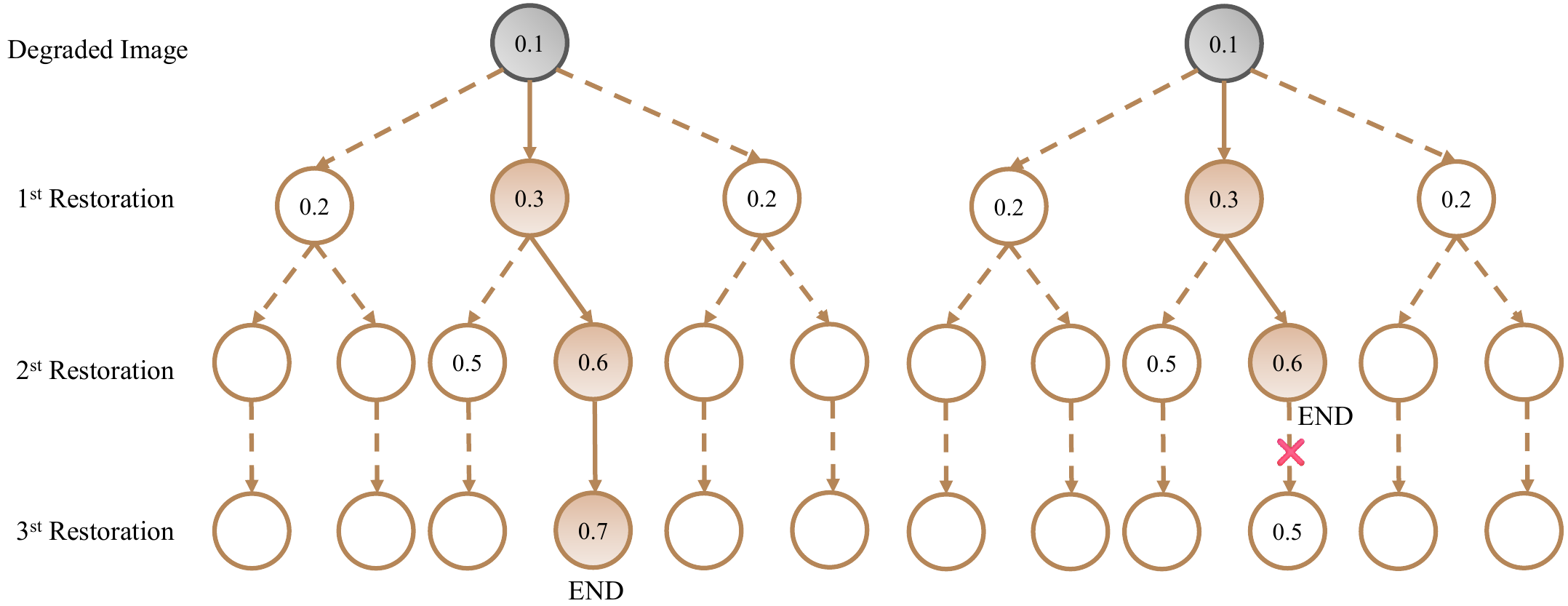}
    \caption{Termination conditions for quality-driven greedy IR strategy.}
    \label{fig:terminate}
    \vspace{-0.6cm}
\end{figure}

Specifically, the method integrates IQA and IR tools. At each step, IR first applies all possible restoration operations to the current image, followed by an evaluation of the restoration results using IQA methods. The restoration operation that yields the highest image quality is selected as the next step, and the current image is updated in preparation for the subsequent iteration. As shown in Fig.~\ref{fig:terminate}, two termination conditions are defined in this approach: 1) All image degradations, as perceived by the MLLM, have been restored by the corresponding IR tools; 2) No further restoration operations in the subsequent iteration lead to an improvement in the image quality. Images that meet either of these conditions are returned to the user.
\subsubsection{Image Quality Guidance}
Objective IQA is a critical component of Q-Agent, as it provides quality feedback on the restored images, thereby guiding the IR process in a systematic and informed manner. Unlike reference-based IQA metrics such as PSNR, SSIM \cite{ssim}, and LPIPS \cite{lpips}, which require ground-truth images as reference, the non-reference (NR) IQA methods are employed in the agent, enabling the prediction and monitoring of image quality throughout the restoration process. 
To ensure a comprehensive and accurate objective IQA, we consider 5 representative NR methods.
Specifically, NIQE \cite{niqe} focuses primarily on the natural scene statistical features of the image, BRISQUE \cite{brisque} integrates both lighting and structural information, and CPBD \cite{cpbd} is widely used for evaluating image quality in the presence of blurring and compression artifacts. Additionally, CLIP-IQA \cite{clipiqa} emphasizes the semantic information of the image, while Hyper-IQA \cite{hyperiqa} is capable of detecting local distortions. 
We denote these metrics as $v=[ni,br,cp,cl,hy]$, respectively.
%
The overall image quality can be defined as the mean of all image quality:
\begin{equation}
\label{eq:quality}
Quality = \frac{{cp + cl + hy - ni - br}}{5}.
\end{equation}
A higher $Quality$ value indicates better image quality.
At each stage of the restoration process, the order with the highest $Quality$ is selected as the optimal operation.

\subsubsection{Autonomous IR Model Selection}
Quality-driven greedy restoration likewise provides an autonomous solution for IR model selection. In each restoration, different IR models are also compared, and the IR model is finally selected with the goal of maximizing the quality gain. After statistical analysis, the various IR models that are used most frequently are outlined in Table~\ref{tab:ir}.
Here, we consider 8 typical image degradations in real-world scenarios: Gaussian noise, JPEG compression artifacts, rain, haze, motion blur, out-of-focus blur, low illumination, and low resolution. 
Several observations can be made from Table~\ref{tab:ir}: 1) For Gaussian noise, we classify the degradation into three categories based on the degree of degradation, and select denoising algorithms with varying levels of noise removal to optimize the restoration; 2) With the exception of DeJPEG and Deraining, Q-Agent makes the most use of Task-Specific models, highlighting the effectiveness of Task-Specific models for specific degradation restoration; 3) All employed IR models are pre-trained on their respective datasets, underscoring the extensibility of the Q-Agent for future expansion and adaptation to new tasks.

\begin{table}[!tp]
    \centering
    \caption{The most commonly used IR tools by Q-Agent. }
    \resizebox{1\linewidth}{!}{\begin{tabular}{c|c|c|c}
    \toprule
         Task & Label & IR Tools & Pretrained Dataset \\    \hline
         \multirow{3}{*}{Gaussian Denoise} &\multirow{3}{*}{DN}& SCUNet (Low Intensity) \cite{scunet} & KodaK24 \cite{kodak}\\
         & & SCUNet (Middle Intensity) \cite{scunet} & KodaK24 \cite{kodak}\\
         & & SCUNet (High Intensity) \cite{scunet} & KodaK24 \cite{kodak}\\ \hline
         DeJPEG & DJ& \multirow{2}{*}{Restormer \cite{Zamir2021Restormer}} & BSDS500 \cite{arbelaez2010contour}\\
         Deraining &DR & &Rain100 \cite{yang2017deep}\\ \hline
         Dehazing &DH& RIDCP \cite{wu2023ridcp} & RIDCP500 \cite{wu2023ridcp} \\
         Motion Deblurring &MDB& AdaRevD \cite{mao2024adarevd} & GoPro \cite{nah2017deep}\\
         Defocus Deblurring &DDB & DRBNet \cite{ruan2022learning} & DPDD \cite{abuolaim2020defocus}\\
         Low-light Enhancement &LE & Retinexformer \cite{Cai_2023_ICCV}& SDSD-indoor \cite{wang2021seeing}\\
         Super Resolution & SR& BSRGAN \cite{zhang2021designing}&RealSRSet \cite{zhang2021designing}\\

    \bottomrule
    \end{tabular}}
    \label{tab:ir}
    \vspace{-0.3cm}
\end{table}

\begin{table*}[]
    \centering
    \caption{Degradation perception accuracy of different MLLMs on the Q-Degrade dataset, where NI-L, NI-M, NI-H, JP, RA, HA, MB, DB, LL, LR denote Noise (Low), Noise (Middle), Noise (High), JPEG Compression Artifact, Rain, Haze, Motion Blur, Out-of-Focus Blur, Low Light, Low Resolution respectively. Best in {\bf\textcolor{red}{RED}}, second in {\bf\textcolor{blue}{BLUE}}.}
    \vspace{-0.3cm}
    \resizebox{\linewidth}{!}{\begin{tabular}{l|cccccccccc|c}
         \toprule
         \multirow{2}{*}{MLLM}  & \multicolumn{10}{c|}{$DACC$} & $MACC$ \\ \cdashline{2-12}
         & NI-L& NI-M& NI-H       & JP      & RA     & HA     & MB& DB      & LL      & LR       & $All$    \\  \hline
         \multicolumn{5}{l}{\textit{\textbf{Zero-shot}}}\\ \hdashline

         GPT-5.4 \cite{gpt5}  & 0.8343  &\bf\textcolor{blue}{0.7651} &0.8704 & 0.7662 & 0.7831 & 0.7639 & 0.8439 & 0.8358 & 0.8135 & 0.8364 & 0.0582 \\
        Gemini-3.1 Pro \cite{gemini3}& 0.8426 & 0.7540 & 0.7842 & 0.5879 & \bf\textcolor{blue}{0.9141} & 0.4239 & 0.8788 & 0.8570 & 0.6239 & 0.5421 & 0.0634\\
         Llava-Llama3-8B (In RestoreAgent) \cite{touvron2023llama,liu2024llavanext}& 0.0030 & 0.0125 & 0.0753 & 0.0131 & 0.7045 & 0.0414 & 0.8238 & 0.8236 & 0.0000 & 0.0000 & 0.0701  \\  
         \textbf{Qwen2-VL-7B-Instruct (In Q-Agent)} \cite{qwen2vl}&0.7542 & 0.7233 & 0.7446 & 0.5056 & 0.8263 & 0.3772 & 0.7675 & 0.7989 & 0.3417 & 0.2049 & 0.1053\\ \hline
         \multicolumn{5}{l}{\textit{\textbf{Zero-shot + CoT}}}\\ \hdashline
         GPT-5.4 \cite{gpt5}  & 0.8582 & 0.7435 & \bf\textcolor{blue}{0.8722} & \bf\textcolor{blue}{0.8340}  &0.7792 & 0.7830 & 0.9035 & 0.9026& 0.7631 & \bf\textcolor{blue}{0.8445}&  0.5939\\
         Gemini-3.1 Pro \cite{gemini3}& 0.8501 & 0.7546 & 0.8621 & 0.7262 & \bf\textcolor{red}{0.9190} & 0.7560 & \bf\textcolor{blue}{0.9081} & 0.9133 & \bf\textcolor{blue}{0.8732} & 0.7594 & 0.5847  \\
         Llava-Llama3-8B (In RestoreAgent) \cite{touvron2023llama,liu2024llavanext}& 0.6323 & 0.5989 & 0.5577 &0.6834 & 0.8144 & 0.5206 & 0.7890 & 0.8389 & 0.5047 & 0.4683 & 0.4779  \\  
         \textbf{Qwen2-VL-7B-Instruct (In Q-Agent)} \cite{qwen2vl}&\bf\textcolor{blue}{0.9097} & 0.7478 & 0.8022 & 0.7978 & 0.8790 & \bf\textcolor{blue}{0.8464} & 0.8829 & \bf\textcolor{blue}{0.9180} & 0.7836 & 0.7771 & \bf\textcolor{blue}{0.6033} \\ \hline
         \multicolumn{5}{l}{\textit{\textbf{Fine-tuned + CoT}}}\\ \hdashline
        \textbf{Qwen2-VL-7B-Instruct (In Q-Agent)}\cite{qwen2vl}&\bf\textcolor{red}{0.9495} & \bf\textcolor{red}{0.9235} & \bf\textcolor{red}{0.9441} & \bf\textcolor{red}{0.8819} & 0.9010 & \bf\textcolor{red}{0.8908}  & \bf\textcolor{red}{0.9430} & \bf\textcolor{red}{0.9208} & \bf\textcolor{red}{0.8875} & \bf\textcolor{red}{0.9588} & \bf\textcolor{red}{0.7882}\\

         \bottomrule
    \end{tabular}}
    \label{tab:mllm}
    \vspace{-0.4cm}
\end{table*}

\section{Experiments}
\subsection{Experimental Setup and Criterion}
In this section, we present a series of comprehensive experiments conducted to evaluate the effectiveness and robustness of the proposed Q-Agent framework. Specifically, the experiments are designed to assess the Q-Agent's degradation perception ability and IR effectiveness. To ensure a fair comparison, the large-scale IR dataset proposed in this paper, Q-Degrade, is divided into a training set and a test set in an 8:2 ratio, with no content overlap between the two sets. The training set is used to fine-tune the Qwen2-VL-7B-Instruct \cite{qwen2vl}, which is utilized in Q-Agent for degradation perceptions, while all experimental evaluations are carried out on the test set. All IR tools employed in Q-Agent are listed in Supplementary Material. The experiments are conducted on a server equipped with 8 × A100 GPUs to facilitate efficient processing.

To assess various aspects of the Q-Agent's performance, a set of distinct metrics has been developed and employed. For MLLM-based degradation perception, we first introduce two vectors: the degradation vector $D$ and the perception vector $\hat D$. Both vectors are 10-dimensional binary vectors, where each element indicates the presence or absence of a specific type of degradation. To evaluate the performance of multiple degradation perception, we define the accuracy metric $MACC$ as follows:
\begin{equation}
   MACC = \frac{T}{{T + F}},
\end{equation}
where $T$ represents the number of instances where the degradation vector $D$ and perception vector $\hat D$ are identical, and $F$ denotes the number of instances where they differ. $MACC$ serves as a global metric, but lacks a targeted examination of different degradation types. For this reason, we introduce the Degradation Accuracy $DACC$, which is adapted from Zhou $et$ $al.$ \cite{zhou2024memo} and defined as follows:
\begin{equation}
   DACC = P({\hat S_i}|{S_i}),i = 1,2,...,10,
\end{equation}
where $S_i$ represents the set of images with the $i$-th type of degradation, and $\hat S_i$ refers to the subset of images detected by the MLLM as exhibiting the $i$-th degradation type. In addition, to evaluate the effectiveness of the quality-driven greedy restoration process, we employ four widely used metrics in IR: PSNR, SSIM \cite{ssim}, LPIPS \cite{lpips}, and DISTS \cite{dists}. Among these, higher values of PSNR and SSIM indicate better IR quality, while LPIPS and DISTS exhibit a negative correlation with restoration effectiveness. The average of the restoration performance of various restoration schemes over all images in the test set is recorded as the final performance.

\begin{figure*}[!t]
    \vspace{-0cm}
    \centering
    \includegraphics[width =1\linewidth]{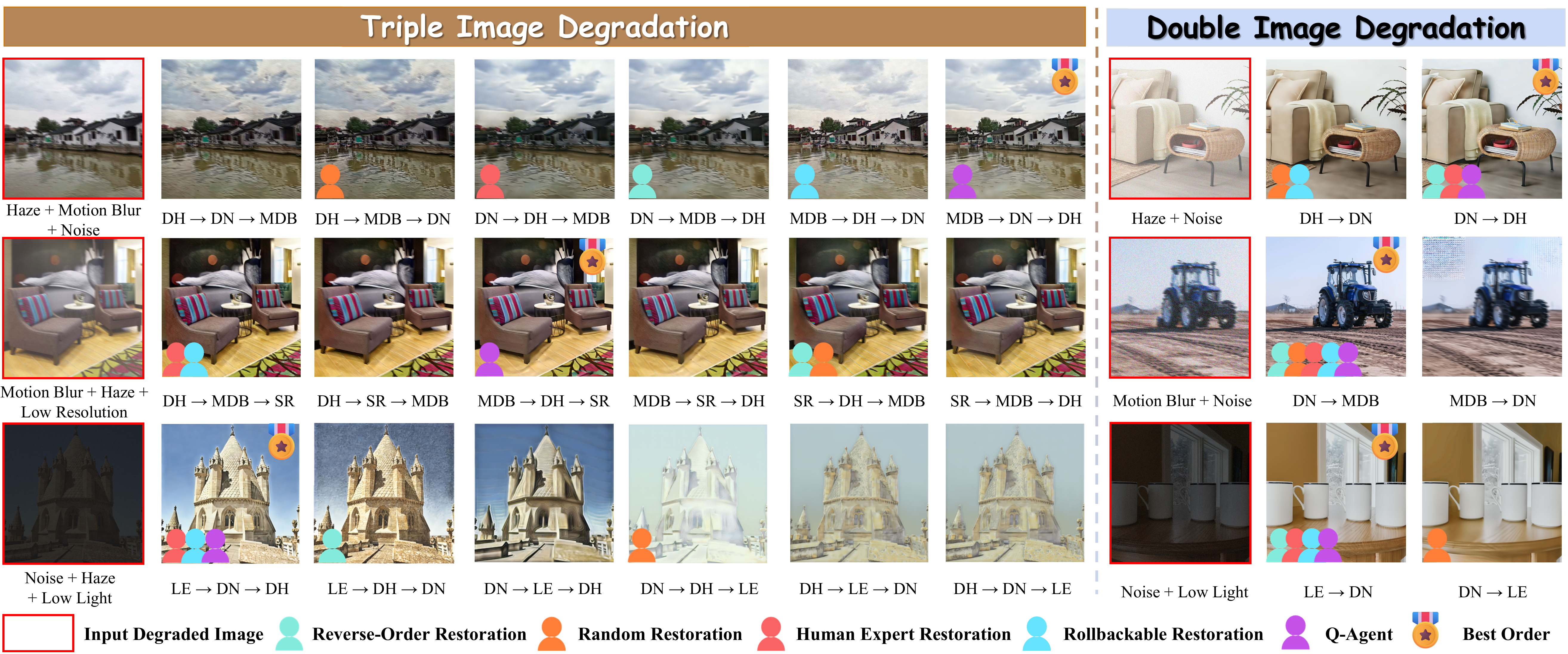}
    \vspace{-0.4cm}
    \caption{Results of different sequential selection strategies. The labeling of the input image contains the degradation type and its order. Notably, the Rollbackable Restoration strategy only introduces the operation of rollback in greedy restoration, and is not exactly the same as the strategy in RestoreAgent \cite{chen2025restoreagent}, due to the fact that the authors have not open-sourced the relevant code and data.}
    \label{fig:deseq}
    \vspace{-0.4cm}
\end{figure*}

\subsection{Accuracy of Degradation Prediction}
To evaluate the performance of MLLMs in perceiving multiple degradation types, we select 3 representative MLLMs for comparison. Among these, GPT-5.4 \cite{gpt5} and Gemini-3.1 Pro \cite{gemini3} are closed-source models, while Llava-Llama3-8B, which is employed in RestoreAgent \cite{chen2025restoreagent}, implemented using open-source code provided by author. The whole experiment is divided into 3 different inference cases to obtain a comprehensive validation and the results are summarized in Table~\ref{tab:mllm}. 

From the results presented in Table~\ref{tab:mllm}, several key observations can be made: 1) In the cases of zero-shot inference, though MLLMs demonstrate satisfactory performance in perceiving individual degradation types, all existing MLLMs exhibit significant limitations in perceiving multiple degradation types; 2) Qwen2-VL \cite{qwen2vl} achieves the best multiple degradation perception performance in all cases, justifying its selection as the foundational model in the Q-Agent framework; 3) By introducing CoT and fine-tuning, except in a few cases, the degradation perception performance of MLLMs improves, especially in multiple degradation awareness, demonstrating the effectiveness of utilized techniques; 4) The Qwen2-VL-7B \cite{qwen2vl} employed in Q-Agent outperforms Llava-Llama3-8B \cite{touvron2023llama,liu2024llavanext} utilized in RestoreAgent \cite{chen2025restoreagent} with smaller size.

\subsection{Order of Restoration Models}
To more effectively and accurately compare the impact of different restoration orders on the degraded images, we design and evaluate 5 distinct restoration strategies: 1) \textbf{Reverse-Order Restoration}: images are restored sequentially in reverse order based on the sequence in which degradation is applied; 2) \textbf{Random Restoration}: the restoration order is randomly selected from the degradation types perceived by the MLLM; 3) \textbf{Human Expert Restoration}: human experts restore the degraded images by directly perceiving the degradation types and determining the restoration order accordingly; 
4) \textbf{Rollbackable Restoration}: A rollbackable search strategy used by RestoreAgent \cite{chen2025restoreagent}; 5) \textbf{Q-Agent Restoration}: a quality-driven greedy strategy is used to determine the restoration order. The results of all strategies are presented in Table~\ref{tab:order} and Fig.~\ref{fig:deseq}. 

From the experimental results, several conclusions can be drawn: 1) Among all restoration strategies, the quality-driven greedy restoration approach employed by Q-Agent achieves the best restoration results, as evidenced by both subjective visual assessments and objective metrics. This superior performance can be attributed to the role of the NR IQA tools, which provides continuous evaluation and guidance throughout the restoration process, making the sequence selection more effective and stable; 2) Although reverse-order restoration lacks practical real-world applications, its performance underscores that IR is not a simple reversible process. This reinforces the critical importance of selecting the appropriate restoration sequence for optimal results; 3) While the rollbackable restoration method employed in RestoreAgent \cite{chen2025restoreagent} is largely able to achieve solutions that exceed that restored by human experts, it can be seen in conjunction with Fig.~\ref{fig:deseq} that in some cases the order of restoration determined by this strategy is still not optimal, mainly due to the lack of quality guidance.

\begin{figure*}[!t]
    \centering
    \includegraphics[width =1\linewidth]{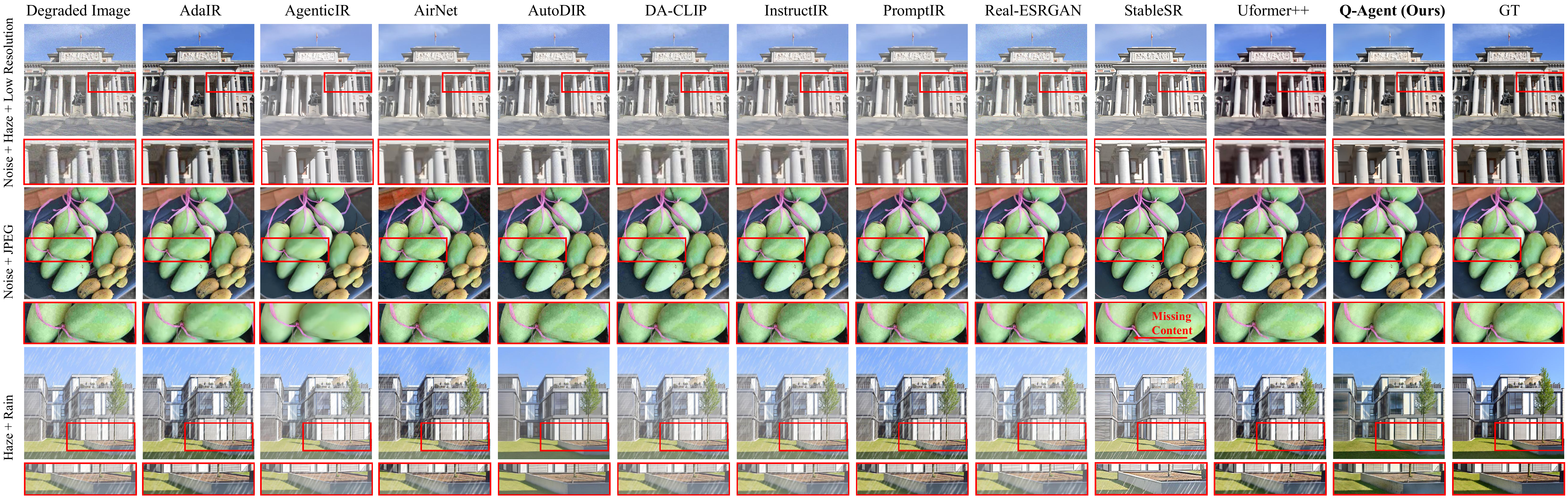}
    \vspace{-0.6cm}
    \caption{Performance comparison of Q-Agent with other All-in-One methods. The red box zooms in locally for a better view.}
    \label{fig:allinone}
    \vspace{-0.3cm}
\end{figure*}

\begin{table}[!t]\centering
\caption{Performance comparison of different restoration order selection strategies. Best in {\bf\textcolor{red}{RED}}, second in {\bf\textcolor{blue}{BLUE}}.}

\label{tab1}
\setlength{\tabcolsep}{1mm}
\resizebox{\linewidth}{!}{\begin{tabular}{l|c|c|c|c}
\toprule
Strategy &  PSNR$\uparrow$ & SSIM$\uparrow$ & LPIPS$\downarrow$ & DISTS$\downarrow$\\
\hline
Reverse-Order Restoration& 19.6072 & 0.5703 & 0.2037 & 0.1017\\
Random Restoration& 18.6369 & 0.5870 & 0.1916 & 0.1044\\
Rollbackable Restoration & \bf\textcolor{blue}{21.0382} & \bf\textcolor{blue}{0.6483} & \bf\textcolor{blue}{0.1840} & 0.0989 \\
Human Expert Restoration& 20.1890 & 0.6288 & 0.2018 & \bf\textcolor{blue}{0.0950}\\ \hline
\textbf{Q-Agent (Ours)}&  \bf\textcolor{red}{22.5847} & \bf\textcolor{red}{0.6856} &\bf\textcolor{red}{0.1529} & \bf\textcolor{red}{0.0778} \\

\bottomrule

\end{tabular}}
\vspace{-0.5cm}
\label{tab:order}
\end{table}

\begin{table}[!t]\centering
\caption{Performance comparison of Q-Agent with different All-in-One models. Best in {\bf\textcolor{red}{RED}}, second in {\bf\textcolor{blue}{BLUE}}.}
\label{tab1}
\setlength{\tabcolsep}{1mm}
\resizebox{0.92\linewidth}{!}{\begin{tabular}{l|c|c|c|c}
\toprule
Method &  PSNR$\uparrow$ & SSIM$\uparrow$ & LPIPS$\downarrow$ & DISTS$\downarrow$\\
\hline
AirNet \cite{airnet}&17.7129 &0.4744 & 0.2531 & 0.2352\\
AutoDIR \cite{autodir}& 18.4029 & 0.4320 & \bf\textcolor{blue}{0.2313} &0.1885\\
DA-CLIP \cite{daclip}& \bf\textcolor{blue}{20.8141} &\bf\textcolor{blue}{0.6471} & 0.2876 & \bf\textcolor{blue}{0.1546} \\
Uformer++ \cite{xu2023uformer++}& 17.7599 & 0.4543 & 0.4534 & 0.2320 \\
InstructIR \cite{insturctir}& 17.5059 & 0.3829 & 0.2595 & 0.2356\\
PromptIR \cite{promptir}& 20.0148 & 0.5259 & 0.2434 & 0.2201\\
Real-ESRGAN \cite{realesrgan}& 13.6988 & 0.2055 & 0.2589 & 0.2600\\
StableSR \cite{stablesr}& 12.4668 & 0.1821 & 0.4235 & 0.2924\\ 
AdaIR \cite{cui2024adair} & 17.5001 & 0.4141 & 0.4506 & 0.1960\\
AgenticIR \cite{zhu2024intelligent}& 19.2197 & 0.5462 & 0.2399 & 0.1494\\ \hline
\textbf{Q-Agent (Ours)}&  \bf\textcolor{red}{22.5847} & \bf\textcolor{red}{0.6856} &\bf\textcolor{red}{0.1529} & \bf\textcolor{red}{0.0778}\\

\bottomrule

\end{tabular}}
\vspace{-0.5cm}
\label{tab:allinone}
\end{table}

\subsection{Comparisons with All-in-One Methods}
From the perspective of both function and structure, the Q-Agent can be regarded as a highly flexible and expandable All-in-One approach. To validate the effectiveness of the Q-Agent, we compare it with 10 representative All-in-One methods, and the experimental results are presented in Table~\ref{tab:allinone}. It is important to note that, to ensure fairness, all selected methods utilize the code and pre-trained models provided by the respective authors. Additionally, to provide a more intuitive representation of the restoration effects achieved by the various All-in-One methods, we have included visualized results, as shown in Fig.~\ref{fig:allinone}. Upon analyzing the results, several key conclusions can be drawn: 1) In terms of restoration quality, Q-Agent outperforms all other All-in-One methods and leads the second IR method by a significant margin of 1.7 dB PSNR, highlighting the effectiveness of the framework; 2) The superior performance of Q-Agent can be attributed to its integration of various Task-Specific IR tools, which enables end-to-end processing while leveraging the effectiveness of specialized restoration methods; 3) Although Q-Agent achieves optimal performance in the current experiment, its performance can be further improved as new, more advanced IR methods are developed, thereby underscoring the extensibility of the proposed framework.
\begin{table}[!t]\centering
\caption{Comparison of degradation perception performance of different strategies. Best in {\bf\textcolor{red}{RED}}, second in {\bf\textcolor{blue}{BLUE}}.}
\label{tab1}
\setlength{\tabcolsep}{1mm}
\resizebox{\linewidth}{!}{\begin{tabular}{lcc}
\toprule
        Method & Accuracy ($MACC$)  \\ \hline
      DepictQA (Zeroshot) \cite{you2024depicting} & 0.5225\\
      DepictQA (Finetuned) \cite{you2024depicting}& \bf\textcolor{blue}{0.7156}\\
      Multihot classifier (Finetuned) &  0.5741\\
      Qwen2-VL (Zeroshot) \cite{qwen2vl}& 0.1053\\ 
      Qwen2-VL (Zeroshot) \cite{qwen2vl} + CoT \cite{wei2022chain} & 0.6033\\ 
      \textbf{Qwen2-VL (Finetuned) + CoT \cite{wei2022chain}} & \bf\textcolor{red}{0.7882}\\

\bottomrule
\end{tabular}}
\vspace{-0.2cm}
\label{tab:degrade}
\end{table}

\begin{table*}[!t]\centering

\caption{Experimental results of different IR agents and strategies on Q-Degrade and DPED datasets. Best in {\bf\textcolor{red}{RED}}, second in {\bf\textcolor{blue}{BLUE}}.}
\vspace{-0.1cm}
\label{tab1}
\setlength{\tabcolsep}{1mm}
\resizebox{\linewidth}{!}{\begin{tabular}{l|c|c|c|c|c|c|c|c|c|c}
\toprule
\multirow{2}{*}{Method} &\multicolumn{5}{c|}{Q-Degrade} & \multicolumn{5}{c}{DPED} \\ \cline{2-11}
 &  PSNR$\uparrow$ & SSIM$\uparrow$ & LPIPS$\downarrow$ & DISTS$\downarrow$ & Run Time (s) &  PSNR$\uparrow$ & SSIM$\uparrow$ & LPIPS$\downarrow$ & DISTS$\downarrow$ & Run Time (s)\\
\hline
AgenticIR \cite{zhu2024intelligent}& 19.2197 & 0.5462 & 0.2399 & 0.1494& 76.92 & 20.0412 &0.9149 & 0.1752 & 0.1185 & 72.41 \\
RestoreAgent (Unofficial) \cite{chen2025restoreagent}& 20.4641 & 0.5883 & 0.2125 & 0.1387 & 93.55 & 21.7522 & 0.9034 & 0.1435 & 0.1084 & 90.65\\ 
Q-Agent (Beam Search)&  \bf\textcolor{red}{23.7089} & \bf\textcolor{red}{0.6941} & \bf\textcolor{red}{0.1505}& \bf\textcolor{red}{0.0766} & \bf\textcolor{blue}{29.41} & \bf\textcolor{red}{25.0672} & \bf\textcolor{red}{0.9621} & \bf\textcolor{red}{0.0794} & \bf\textcolor{red}{0.0607} & \bf\textcolor{blue}{22.36}\\
\textbf{Q-Agent (Greedy Search)}&  \bf\textcolor{blue}{22.5847} & \bf\textcolor{blue}{0.6856} &\bf\textcolor{blue}{0.1529} & \bf\textcolor{blue}{0.0778}& \bf\textcolor{red}{21.24}  &  \bf\textcolor{blue}{23.5071} & \bf\textcolor{blue}{0.9544} & \bf\textcolor{blue}{0.0916} & \bf\textcolor{blue}{0.0633} & \bf\textcolor{red}{15.13}\\

\bottomrule

\end{tabular}}
\vspace{-0.5cm}
\label{tab:addexp}
\end{table*}

\subsection{Comparison of Algorithm Generalizability}
To evaluate the effectiveness of the proposed Q-Agent in restoring real-world degraded images, we conduct experiments using the well-established DSLR Photo Enhancement Dataset (DPED) \cite{ignatov2017dslr}. This dataset comprises over 22,000 images captured by various smartphones and DSLR cameras, encompassing a wide range of authentic image degradations. As shown in Table~\ref{tab:addexp}, the Q-Agent consistently outperforms both the All-in-One methods and the IR Agents across all metrics, thereby demonstrating its superior capability in addressing real-world image degradation. This performance advantage is largely attributed to the integration of perceptual quality metrics (such as NIQE \cite{niqe} and BRISQUE \cite{brisque}) within quality guidance, which enhances agent's sensitivity to authentic degradations.

\subsection{Validity of CoT}
The use of CoT \cite{wei2022chain} requires multi-step reasoning, particularly in complex or unknown degradation scenarios, since it can explicitly reason about the presence or absence of each degradation type, improving recognition of co-occurring degradations.
To validate the contribution of CoT, we employ DepictQA \cite{you2024depicting} and implement a multi-hot classifier as a baseline. 
As shown in Table~\ref{tab:degrade}, CoT significantly enhances the accuracy of degradation perception.
In contrast, multi-hot classifier suffer from severely impaired restoration quality due to incomplete or imprecise perception of degradation types, particularly in out-of-domain cases (where there may exist a degradation not in the training set).

\subsection{Quality Metric Ablation}
To validate the effectiveness and rationality of the employed quality metric $Quality$, which is defined in Equ.~\ref{eq:quality}, we conduct ablation experiments focusing on different components. Specifically, we first carry out a subjective quality assessment study in which multiple human participants evaluate images from the Q-Degrade dataset. Subsequently, we apply several NR IQA methods to predict the quality scores of these images. The alignment between the NR IQA metrics and human visual perception is quantified using 4 widely adopted statistical measures: Spearman Rank Correlation Coefficient (SRCC) and Pearson Linear Correlation Coefficient (PLCC), Kendall Rank-Order Correlation Coefficient (KRCC), and Root Mean Square Error (RMSE). The results, summarized in Table \ref{tab:iqaablation}, demonstrate that the proposed $Quality$ metric, formed by integrating multiple NR IQA methods, achieves a high degree of consistency with human judgments. Notably, the weights derived through support vector regression (SVR) for the five combined NR IQA methods closely approximate an equal weighting scheme. Based on this observation, we adopt an average weighting strategy for the $Quality$ metric employed by the Q-Agent. By further analyzing the results, we believe that the effectiveness of this average combination strategy is due to the fact that different NR IQA methods focus on different dimensions of image quality. Specifically, NIQE \cite{niqe} is employed to characterize the statistical regularities of natural scenes, while CPBD \cite{cpbd} primarily targets the detection of image blurriness. BRISQUE \cite{brisque} assesses image quality by jointly considering luminance and structural information. CLIP-IQA \cite{clipiqa} emphasizes semantic-level features, whereas Hyper-IQA \cite{hyperiqa} is effective in identifying localized distortions.
\begin{table}[!t]\centering
\caption{Ablation Results for NR IQA methods on Q-Degrade. Best in {\bf\textcolor{red}{RED}}, second in {\bf\textcolor{blue}{BLUE}}.}
\label{tab1}
\setlength{\tabcolsep}{1mm}
\resizebox{0.88\linewidth}{!}{\begin{tabular}{lcccc}
\toprule
Method &  SRCC$\uparrow$ & PLCC$\uparrow$ & KRCC$\uparrow$ & RMSE$\downarrow$\\
\hline
NIQE \cite{niqe}& 0.4219 &0.4516 & 0.3125 & 0.8781\\
CPBD \cite{cpbd}& 0.4695 & 0.4871 & 0.3326 & 0.8544\\
BRISQUE \cite{brisque}& 0.5765 & 0.5938 & 0.4371 &0.7862\\
CLIP-IQA \cite{clipiqa} & \bf\textcolor{blue}{0.7251} & \bf\textcolor{blue}{0.7412} & \bf\textcolor{blue}{0.5219} & \bf\textcolor{blue}{0.5590}\\
Hyper-IQA \cite{hyperiqa}& 0.6233 & 0.6770 & 0.4710 & 0.7538 \\ \hline
\textbf{All (SVR)} & \bf\textcolor{red}{0.8122} & \bf\textcolor{red}{0.8351} & \bf\textcolor{red}{0.6231} & \bf\textcolor{red}{0.4835}\\

\bottomrule

\end{tabular}}
\vspace{-0.4cm}
\label{tab:iqaablation}
\end{table}

\subsection{Dataset Ablation}
To evaluate the robustness of the Q-Agent, we conduct a series of ablation experiments. Specifically, we examine how variations in the size of the training dataset impact the multiple degradation perception and IR capabilities of the Q-Agent framework. The experimental settings and results are presented in Table~\ref{tab:abl}. Upon analyzing the results in Table~\ref{tab:abl}, the following observations can be made: 1) The Q-Agent maintains competitive performance even when fine-tuned on a smaller dataset, demonstrating the framework’s robustness; 2) Overall, as $MACC$ decreases, there is a corresponding decline in the quality of IR, suggesting that the performance of the Q-Agent is influenced by the MLLM's ability to sense multiple degradations.
\begin{table}[!t]\centering

\caption{Results of Q-Agent ablation experiments on the Q-Degrade dataset. Best in {\bf\textcolor{red}{RED}}, second in {\bf\textcolor{blue}{BLUE}}.}
\label{tab1}
\setlength{\tabcolsep}{1mm}
\resizebox{\linewidth}{!}{\begin{tabular}{c|c|c|c|c|c}
\toprule
Train / Test Scale &$MACC$$\uparrow$&  PSNR$\uparrow$ & SSIM$\uparrow$ & LPIPS$\downarrow$ & DISTS$\downarrow$\\
\hline
20K / 20K& 0.7092& 19.7588 & 0.6022 & 0.2131 & 0.1435\\
40K / 20K& 0.7596& 20.2483 & 0.6581 & 0.1753 & 0.0944\\
60K / 20K& \bf\textcolor{blue}{0.7723}& \bf\textcolor{blue}{21.9045} & \bf\textcolor{blue}{0.6798} & \bf\textcolor{red}{0.1462} & \bf\textcolor{blue}{0.0786}\\ 
80K / 20K& \bf\textcolor{red}{0.7882}&  \bf\textcolor{red}{22.5847} & \bf\textcolor{red}{0.6856} &\bf\textcolor{blue}{0.1529} & \bf\textcolor{red}{0.0778} \\

\bottomrule

\end{tabular}}
\vspace{-0.4cm}
\label{tab:abl}
\end{table}

\section{Limitations \& Potential Improvement}
While the effectiveness and robustness of the proposed Q-Agent have been validated from multiple perspectives, an inherent limitation remains in its reliance on a greedy search strategy, which may lead to suboptimal restoration results due to convergence to local optima. To address this issue, we explore the use of a beam search strategy to enhance the decision-making process. Specifically, at each stage of the degradation restoration process, the Q-Agent retains the top $K$ candidate actions associated with the highest $Quality$ scores. To evaluate the impact of this modification, we set $K=2$ and conduct experiments on both the Q-Degrade and DPED \cite{ignatov2017dslr} datasets. As illustrated in Tabel~\ref{tab:addexp}, the beam search strategy achieves improved restoration performance compared to the greedy search approach. However, this enhancement comes at the cost of increased computational complexity and runtime. Consequently, we suggest that optimizing the trade-off between computational efficiency and restoration quality remains a critical consideration for the design of intelligent restoration agents.

\section{Conclusion}
The effectiveness and generalizability of image restoration (IR) methods have been a subject of extensive research and discussion. The emergence of MLLMs introduces a novel approach to advancing IR techniques. To achieve a balance between effectiveness, versatility, usability, and scalability, we propose Q-Agent, a quality-driven Chain-of-Though (CoT) IR agent. The Q-Agent consists of two key stages: robust degradation perception and quality-driven greedy restoration. Specifically, the CoT reasoning approach decomposes multiple degradation perception task into single degradation perceptions to improve the degradation prediction accuracy of MLLMs. Subsequently, the quality-driven greedy restoration module determines the optimal restoration sequence and applies IR tools guided by no-reference image quality metrics. Experimental results demonstrate that Q-Agent offers significant advantages in terms of degradation perception, IR performance and robustness. As technologies continue to evolve, the Q-Agent is expected to enhance its IR capabilities further, providing a more comprehensive and effective solution for applications.

\ifpeerreview \else

\bibliographystyle{IEEEtran}
\bibliography{references}

\ifpeerreview \else






\fi

\end{document}